\documentclass[acmsmall,natbib=false,authorversion,nonacm]{acmart}

\usepackage[x11names,cmyk]{xcolor}
\usepackage{soul}
\usepackage{listings}
\usepackage{lstautogobble}
\usepackage{lstlinebgrd}
\usepackage{pgffor}

\usepackage{import}
\usepackage{pdfpages}
\usepackage{transparent}

\setcopyright{acmlicensed}
\copyrightyear{2018}
\acmYear{2018}
\acmDOI{XXXXXXX.XXXXXXX}

\acmJournal{JACM}
\acmVolume{37}
\acmNumber{4}
\acmArticle{111}
\acmMonth{8}

\RequirePackage[
  datamodel=acmdatamodel,
  style=acmauthoryear,
  ]{biblatex}

\begin{document}

\title{Reactive Graphs for Efficient Markov Chain Monte Carlo Inference in Probabilistic Programming Languages}

\author{Viktor Palmkvist}
\email{vipa@kth.se}
\affiliation{
  \institution{KTH Royal Institute of Technology}
  \city{Stockholm}
  \country{Sweden}
}
\orcid{0000-0003-0669-4085}

\author{Fredrik Ronquist}
\email{fredrik.ronquist@nrm.se}
\affiliation{
  \institution{Swedish Museum of Natural History}
  \city{Stockholm}
  \country{Sweden}
}
\orcid{0000-0002-3929-251X}

\author{David Broman}
\email{dbro@kth.se}
\affiliation{
  \institution{KTH Royal Institute of Technology}
  \city{Stockholm}
  \country{Sweden}
}
\orcid{0000-0001-8457-4105}

\newenvironment{todoenv}
  {\begin{quote} \color{green}\textbf{Author's Note:}}
  {\end{quote}\noindent}
\newcommand{\todo}[1]{\color{green}\textbf{Author's Note:} #1}

\newenvironment{todo-outline}
  {\begin{itemize}\renewcommand{\labelitemi}{TODO}}
  {\end{itemize}}

\newcommand{\sectionBreakMaybe}{}

\newcommand{\incfig}[2][1]{%
    \import{./}{#2.pdf_tex}
}

\pdfsuppresswarningpagegroup=1

\lstdefinestyle{error}{%
  moredelim=**[is][\color{red}]{@}{@},
}

\definecolor{ckeywords}{rgb}{0.13,0.13,1}
\definecolor{ccomments}{rgb}{0,0.5,0.5}
\definecolor{cstrings}{rgb}{0,0.5,0}
\definecolor{cwarnings}{rgb}{1,0.5,0}
\definecolor{codemargin}{rgb}{0.9,0.9,0.9}
\definecolor{listinggray}{rgb}{0.95,0.95,0.95}

\lstset{%
  xleftmargin=2em,
  numbers=left,
  frame=l,
  backgroundcolor=\color{listinggray},
  fillcolor=\color{codemargin},
  framesep=1.3em,
  framexleftmargin=1.0em,
  numberstyle=\normalfont\tiny,
  framerule=0pt,
  lineskip=-1pt,
  basicstyle=\setlength{\lineskip}{-0.1pt}\ttfamily\selectfont\footnotesize
}

\lstdefinelanguage{MCore}{%
    escapechar=\$,
    morekeywords={Lam,con,else,end,fix,if,in,lam,lang,let,match,recursive,sem,syn,then,type,use,utest,with,letop,letimpl,all,repr},
    otherkeywords={->,_,@,?},
    keywordstyle=\color{ckeywords},
    morekeywords=[2]{mexpr,include,never},
    keywordstyle=[2]\color{cwarnings},
    morecomment=[l][\color{ccomments}]{--},
    morestring=[b]",
    stringstyle=\color{cstrings},
    sensitive=true,
    breaklines=true,
    escapeinside={(*@}{@*)},
    stepnumber=1,
    columns=fixed, 
    showstringspaces=false,
    mathescape=true,
    breaklines=true,
    breakatwhitespace=true,
    mathescape=true,
    showstringspaces=false
  }

\lstdefinelanguage{TreePPL}{%
    escapechar=\$,
    morekeywords={model,function,assume,let,for,in,return},
    otherkeywords={->,_,@,?,~},
    keywordstyle=\color{ckeywords},
    morekeywords=[2]{mexpr,include,never},
    keywordstyle=[2]\color{cwarnings},
    morecomment=[l][\color{ccomments}]{--},
    morestring=[b]",
    stringstyle=\color{cstrings},
    sensitive=true,
    breaklines=true,
    escapeinside={(*@}{@*)},
    stepnumber=1,
    columns=fixed, 
    showstringspaces=false,
    mathescape=true,
    breaklines=true,
    breakatwhitespace=true,
    mathescape=true,
    showstringspaces=false
  }

\lstset{
  emph={pure,map,apply,P,assume,weight,join,sub},
  emphstyle={\color{cstrings}},
}

\makeatletter

\newcount\bt@rangea
\newcount\bt@rangeb

\newcommand\btIfInRange[2]{%
    \global\let\bt@inrange\@secondoftwo%
    \edef\bt@rangelist{#2}%
    \foreach \range in \bt@rangelist {%
        \afterassignment\bt@getrangeb%
        \bt@rangea=0\range\relax%
        \pgfmathtruncatemacro\result{ ( #1 >= \bt@rangea) && (#1 <= \bt@rangeb) }%
        \ifnum\result=1\relax%
            \breakforeach%
            \global\let\bt@inrange\@firstoftwo%
        \fi%
    }%
    \bt@inrange%
}
\newcommand\bt@getrangeb{%
    \@ifnextchar\relax%
        {\bt@rangeb=\bt@rangea}%
        {\@getrangeb}%
}
\def\@getrangeb-#1\relax{%
    \ifx\relax#1\relax%
        \bt@rangeb=100000
    \else%
        \bt@rangeb=#1\relax%
    \fi%
}

\newcommand{\highlightRange}[1]{%
  \btIfInRange{\value{lstnumber}}{#1}{\color{orange!30}}{}%
}%
\makeatother

\newcommand{\fig}[1]{Fig.~\ref{#1}}

\newcommand{\selectcodefontsize}{}
\newcommand{\basecode}[1]{%
  \begingroup%
  \sethlcolor{listinggray}%
  \texttt{\selectcodefontsize\hl{#1}}%
  \endgroup%
}
\newcommand{\code}[1]{\ifmmode\mathord{\basecode{#1}}\else\basecode{#1}\fi}

\pdfstringdefDisableCommands{%
  \def\code#1{#1}%
}

\begin{abstract}
  An important aspect of making inference based on a probabilistic
  program practical is efficiency; faster evaluation enables more work
  per unit of time, which can be translated into more
  precision. Inference via Markov chain Monte Carlo has a property
  that can be favorably exploited for efficiency: most proposed
  samples are computed as minor variations of previous samples, i.e.,
  a clever implementation can skip computations pertaining to what is
  unchanged. This paper provides an approach for automatically
  translating a probabilistic program to a dynamic graph, reminiscent
  of functional reactive programming, that explicitly represents data
  dependencies, enabling proposals to only recompute the parts of the
  graph that depend on redrawn random variables. The graph-building
  interface follows familiar functional programming interfaces, which
  also connect to their expressiveness in terms of probabilistic
  programming: models using the applicative functor portion express
  Bayesian networks, while those using monads represent universal
  probabilistic programming languages.
\end{abstract}


\maketitle

\section{Introduction}

Many scientific disciplines involve the analysis of phenomena that are
either random in nature, or can favorably be modelled as random
processes. Examples include
phylogenetics~\cite{yangComputationalMolecularEvolution2006},
economics~\cite{rayBayesianModelBehaviour2008}, and
epidemiology~\cite{duarteProbabilisticEpidemiologicalModel2023}. Some
of these problems can be modelled and solved analytically, giving
exact results, but many cannot; they require sophisticated inference
methods to provide sufficiently accurate approximate
results. Fortunately, much work has been done on such inference
methods, e.g., sequential Monte Carlo
(SMC)~\cite{smithSequentialMonteCarlo2013}, Markov chain Monte Carlo
(MCMC)~\cite{gilksMarkovChainMonte1995}, and variational inference
(VI)~\cite{wainwrightGraphicalModelsExponential2008}.

Each of these present different trade-offs, both theoretically and
practically. For example, SMC and MCMC are precise at the limit (i.e.,
given infinite computation time) while VI can only choose the best fit
amongst a user-supplied family of distributions, whether that is
accurate or not, though it often runs faster than the other two. SMC
and MCMC are thus often used when the target distribution is unknown,
or too difficult to express in the distributional form required by
VI. For Bayesian phylogenetic inference in particular, which motivates
our work, SMC has never gained much
traction~\cite{strumbeljPresentFutureSoftware2024}. Instead, MCMC or
one of its many
variations~\cite{horowitzGeneralizedGuidedMonte1991,homanNoUturnSamplerAdaptively2014,syedNonreversibleParallelTempering2022}
tends to be the inference method of choice.

Efficient inference depends not only on the inference method used, but
also on the means by which the probabilistic model is expressed, along
with how the inference method is implemented. A \emph{probabilistic
programming language} (PPL) lets a user describe a probabilistic model
in the form of a
program~\cite{kozenSemanticsProbabilisticPrograms1979}. Such a program
can a) sample from simpler distributions, including distributions
depending on previously sampled values, b) condition values, for
example on observed data, and c) export values to be inferred. PPLs
inherit the expressiveness and flexibility of programming languages,
but also the difficulty of analysis. This is especially true if the
PPL is universal. We refer to a PPL as \emph{universal} if it supports
models with a statically unbounded number of random variables. This
stands in contrast with classical Bayesian networks, whose models have
statically fixed numbers of random variables.

This paper focuses on efficiently supporting Markov chain Monte Carlo
(MCMC) inference using the Metropolis-Hastings
algorithm~\cite{metropolisEquationStateCalculations1953,hastingsMonteCarloSampling1970}
on models written in a universal PPL. In brief, inference produces one
sample at a time by drawing a proposal and then randomly accepting or
rejecting it based on its likelihood relative to the previous
sample. If the proposal is accepted, then that is the next sample,
otherwise we repeat the previous sample. In a PPL we typically produce
the proposal by randomly redrawing a subset of the random variables
present in the model. This has an important performance implication:
we should only need to recompute the parts of the model that depend on
the redrawn random variables; everything else can be reused as-is.

Of course, this is not a new observation. For example, approaches
using more restrictive modelling approaches, e.g., the graphical
models of RevBayes~\cite{hohnaProbabilisticGraphicalModel2014}, tend
to build graph-representations of their models that explicitly
represent such data-dependencies, and Bayesian networks are graph-like
by definition. In the space of PPLs,
C3~\cite{ritchieC3LightweightIncrementalized2016} uses continuation
passing style and function caching to start execution at changed
random variables and skip unchanged computations,
respectively. Finally, Bali-Phy~\cite{redelingsBAliPhyVersion32021}
has a new
release\footnote{\url{https://github.com/bredelings/BAli-Phy/releases/tag/4.1}}
which appears to similarly build a graph for a more expressive PPL,
though the details are not immediately obvious from the
implementation.

In this paper we solve this problem in a principled way by connecting
the well-explored concept of \emph{functional reactive programming}
(FRP)~\cite{elliottFunctionalReactiveAnimation1997} to PPLs. FRP
centers around graphs of dependent values, where changes in
dependencies are propagated to dependent values automatically, along
with \emph{sources} that can trigger changes and \emph{sinks} that do
not propagate changes. Broadly speaking, our approach can be
summarized in three points:

\begin{itemize}
\item Our graphs have random variables as the only kind of source
  available, propagating a concrete sample. At the same time, each
  random variable is a sink, consuming the distribution of the random
  variable (which affects the likelihood of the proposal) without
  necessarily updating the sample drawn from it.
\item Likelihood is updated by dedicated sink nodes, and tracked as
  hidden state inside the graph.
\item We add support for the rejection of a proposal by adding a
  transactional interface; changes caused by redrawing a set of random
  variables can be rolled back.
\end{itemize}

A subtle but particularly important insight is that a graph represents
a single run of a model, \emph{not} the model itself. This means that
even if the model contains recursion, the graph remains
acyclic. Instead, certain nodes will contain ``sub-graphs''
representing the actual choices made in the sample, roughly equivalent
with the execution trace of an ordinary program. However, each node
contains enough information such that the graph can be modified to
represent any possible sample from the model, e.g., by storing the
recursive functions present in the original model, but the graph
itself remains finite.

Similar to FRP, the core of the graph-building interface forms a
\code{Monad}, with two additional probabilistic constructs:
\code{assume} for introducing new random variables and \code{weight}
for conditioning. The standard \code{Functor}, \code{Applicative}, and
\code{Monad} interfaces also have a nice connection to expressivity:
models using \code{Functor} and \code{Applicative} are as expressive
as Bayesian networks, while \code{Monad} brings universality.

In this paper we make the following contributions:

\begin{itemize}
\item An overarching approach for using a modified version of FRP for
  efficient MCMC inference (Section~\ref{sec:approach}).
\item An interface for manually building graphs and using them to
  write model-specific inference
  (Section~\ref{sec:interface}).
\item An implementation of the aforementioned interface using internal
  mutability (Section~\ref{sec:implementation}).
\item A sketch of an automatic translation from
  TreePPL~\cite{senderovTreePPLUniversalProbabilistic2023}, a
  high-level universal probabilistic programming language, to a
  graph-building program (Section~\ref{sec:automatic-translation}).
\end{itemize}

\sectionBreakMaybe
\section{Overarching Approach and Overview}\label{sec:approach}

\begin{figure}
    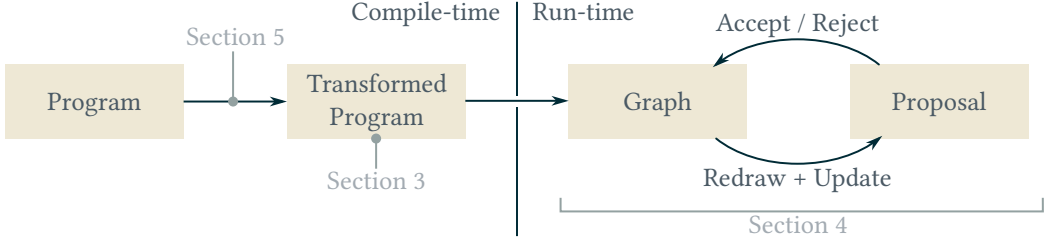

  \caption{The transformations and phases relevant to our approach. We start with a probabilistic program, transform it to a graph-building program, then run it to obtain a graph representation of a sample. Next is inference, which is a cycle of redrawing random variables to create proposals, and then accepting or rejecting them.\label{fig:phases}}
\end{figure}

This section gives an overview of our approach and its phases,
exemplifying each with a single running example. \fig{fig:phases}
shows our phases and artifacts, as well as the structure of this
section.

We begin with a probabilistic \emph{program} written in TreePPL. Our
running example, on the left-hand side of \fig{fig:running-example},
is a common introductory example in Bayesian statistics: inferring the
bias of a coin given a series of observed coin flips. We model the
bias of the coin as a random variable \code{p} with prior distribution
of \code{Beta 1.0 1.0} (line~\ref{line:running-example:prior}), then
loop through our observations (line~\ref{line:running-example:loop}),
observing each in turn as drawn from a \code{Bernoulli} distribution,
parameterized by \code{p}
(line~\ref{line:running-example:observation}), before finally
returning \code{p} (line~\ref{line:running-example:return}). This is
the program a user writes, and contains no mentions or references to
the graph central to our approach; it is an implementation detail.

\begin{figure}
  \begin{minipage}[t]{0.4\linewidth}
    \begin{lstlisting}[language=TreePPL,autogobble=true]
      model function coin
        (flips: Bool[]) => Real
      {
        assume p ~ Beta(1.0, 1.0);$\label{line:running-example:prior}$
        for f in flips {$\label{line:running-example:loop}$
          observe f ~ Bernoulli(p);$\label{line:running-example:observation}$
        }
        return p;$\label{line:running-example:return}$
      }
    \end{lstlisting}
  \end{minipage}
  \hfill
  \begin{minipage}[t]{0.59\linewidth}
    \begin{lstlisting}[language=MCore,autogobble=true]
      let coin = lam flips. lam st.
        match pure (Beta 1.0 1.0) st
          with (st, dist) in
        match assume dist st with (st, p) in
        let loopBody = lam st. lam f.
          let observe =
            lam p. logObserve f (Bernoulli p) in
          match map observe p st with (st, w) in
          (weight w st).0 in
        (foldl loopBody st flips, p)
    \end{lstlisting}
  \end{minipage}
  \caption{Our running example program, before (in TreePPL) and after (in MCore) transformation. The program takes a sequence of observations (coin flips) and describes a posterior distribution for the bias of the coin, \code{p}.\label{fig:running-example}}
\end{figure}

Next, we use a compiler pass to automatically produce a
\emph{transformed program} expressing the same model, but this time
using our graph abstraction. Our approach is implemented in the Miking
framework~\cite{bromanVisionMikingInteractive2019}, whereby the
transformed program is in its OCaml-like core language MCore. A
slightly simplified version of such a program can be seen to the right
in \fig{fig:running-example}. We describe a similar program in more
detail in Section~\ref{sec:interface}, for now it suffices to note the
use of four graph-building operations (\code{pure}, \code{assume},
\code{map}, and \code{weight}), as well as a state variable
(\code{st}) threaded through the program. Note also that the automated
translation, though first in \fig{fig:phases}, is discussed last in
the paper, in Section~\ref{sec:automatic-translation}; its
presentation benefits from the information presented in
Sections~\ref{sec:interface} and~\ref{sec:implementation}.

The transformed program can then be compiled, e.g., together with some
program-agnostic inference code implementing basic MCMC.

\begin{figure}
    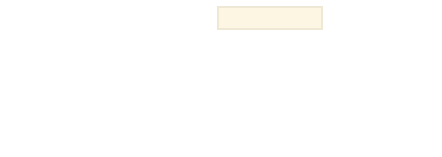

  \qquad
    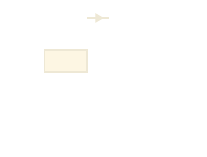

  \caption{A graph (left) produced by running the program in \fig{fig:running-example} and the same graph after producing a new proposal (right). Each node is annotated with the kind of operation that created it. Each proposal mutates the graph, storing new values in some nodes, and produces a reset function that can be called to undo the changes made.\label{fig:graph}}
\end{figure}

Next, running the compiled program produces a \emph{graph}
representing a single sample from the described model. One example,
using the input \code{flips = [true, false, true]}, can be seen on the
left of \fig{fig:graph}. For this particular model, the topology of
the graph will never change; we return to a more complicated example
in Section~\ref{sec:interface} where the topology \emph{does}
change. Note that the graph has no remnant of the looping construct of
the original model, the \code{foldl} of the transformed program has
disappeared. This is a typical result; the \code{flips} is
deterministic, therefore the iteration is deterministic, even if part
of the work done in each iteration involves probabilistic values.

Once this initial graph has been constructed we can start with
inference, which involves computing a new \emph{proposal}, seen on the
right of \fig{fig:graph}. Note that we omit the code content of the
\code{pure} and \code{map} nodes, since they are identical to the
original graph. The inference code picks a random variable (i.e., an
\code{assume}) to redraw, then we propagate updates throughout the
graph. Of course, this simple example has only one random variable,
and updating it updates most of the graph, saving little. Fortunately,
most practically interesting models have more complicated and limited
dependencies, giving ample opportunities for savings.

Our implementation is mutable, i.e., we mutate the original graph to
obtain the proposal. If the proposal is to be rejected we must thus
undo all changes. This is done by making the update function of each
node record the previous value of each internal variable that is
changed, storing them in a ``reset'' function.

We are now ready to describe the graph-building interface in more
detail.

\sectionBreakMaybe
\section{Graph-Building Interface}\label{sec:interface}

The core of the graph-building interface, the most important
operations appearing inside each transformed program, can be seen in
\fig{fig:interface}. In this formulation of the interface the
creation of a node is a stateful action\footnote{It seems plausible to
create a stateless interface, as long as the implementation language
supports observable sharing, except \code{weight}, which would need a
different encoding. However, there are non-obvious tradeoffs involved,
thus we leave this for future work.}, thus each operation returns a
value wrapped in the state monad \code{S}.

\begin{figure}
  \begin{lstlisting}[language=MCore,autogobble=true]
    type S a = State -> (State, a)
    type P a

    let pure   : all a.                        a -> S (P a) $\label{line:interface:pure}$
    let map    : all a. all b.   (a -> b) -> P a -> S (P b)
    let apply  : all a. all b. P (a -> b) -> P a -> S (P b) $\label{line:interface:apply}$

    let assume : all a. P (Dist a) -> S (P a) $\label{line:interface:assume}$
    let weight :           P Float -> S () $\label{line:interface:weight}$

    let join : all a. P (P a) -> S (P a) $\label{line:interface:join}$
    let sub  : all a. P (S a) -> S (P a) $\label{line:interface:sub}$
  \end{lstlisting}
  \caption{The core interface for building a graph. Lines~\ref{line:interface:pure}-\ref{line:interface:weight} form an \code{Applicative Functor} and express Bayesian Networks; adding lines~\ref{line:interface:join} (\code{Monad}) and~\ref{line:interface:sub} gives universality.\label{fig:interface}}
\end{figure}

The core of the interface is the polymorphic type \code{P a}, which
represents a graph node containing a value of type \code{a}. Each
operation creates a new node, potentially with data dependencies on
previous nodes. For example, \code{map} takes a function \code{a -> b}
and a node \code{P a} and creates a new node \code{P b}, a
deterministic transformation of its input node. Together, \code{pure},
\code{map}, and \code{apply} make \code{P} an applicative functor, so
long as we ignore the state monad \code{S}\footnote{These operations
fulfill the \code{Applicative} laws observationally; expressions that
should be equal are observably indistinguishable, even if they
generate graphs with different internal structure.}.

There are two probabilistic nodes: \code{assume}, which introduces new
random variables, and \code{weight}, which imparts a likelihood to the
current sample. Note that the distribution passed to \code{assume} is
itself in a node; this allows random variables parameterized by other
random variables. The \code{weight} node expects a \code{Float}
representing a probability to multiply\footnote{For floating point
precision reasons the implementation uses log-prob numbers, i.e., the
probability is \emph{added} into the likelihood; our explanations omit
this detail in the interest of simplicity.} into the current likelihood.

These five operations are enough to express Bayesian networks. A
(contrived) example of this can be seen in
figures~\ref{fig:running-example:code} (code)
and~\ref{fig:running-example:graph} (graph). In this model, we first
draw parameters
(lines~\ref{line:running-example:prior-start}-\ref{line:running-example:prior-end})
for a Gaussian
(lines~\ref{line:running-example:gaussian-start}-\ref{line:running-example:gaussian-end}). Then,
for each input (line~\ref{line:running-example:for-each}) we draw a
value \code{center} from that Gaussian
(line~\ref{line:running-example:gaussian-draw}) and weight the program
by the probability of drawing the input from a Gaussian with mean
\code{center}
(lines~\ref{line:running-example:observe-start}-\ref{line:running-example:observe-end}). Note
that the distributions describing the first two parameters are wrapped
with \code{pure} to make the types line up, and that we use a
combination of \code{map} and \code{apply} to apply the arity-2
function \code{Gaussian} via currying. Note also that the code
processing the \code{inputs} does so by side-effect (it is only
relevant for computing the likelihood of the sample) and that the
actual iteration takes place \emph{outside} \code{P} values. The
latter point is because the \code{inputs} are given and thus
deterministic, i.e., not wrapped in \code{P}, thus we can do the
iteration once, and then never repeat it.

\fig{fig:running-example:graph} shows the graph produced by running
the code in \fig{fig:running-example:code} with the input \code{[9.1,
    3.4]}. Note that the list itself does not show up, only the
elements themselves, captured by closure in the two \code{map} lambdas
to the right in the figure (originally produced on
line~\ref{line:running-example:observe-start} in
\fig{fig:running-example:code}).

Next, consider how redrawing individual \code{assume}s in
\fig{fig:running-example:graph} affects the graph. For example,
redrawing \code{assume} 3 requires updating the top-right \code{map}
and \code{weight} nodes, but nothing else. Redrawing \code{assume} 1
similarly requires recomputing the \code{Gaussian} \code{map} node and
the following \code{apply} node, but it also requires some likelihood
updates in~3 and~4. Those \code{assume}s will not be redrawn, thus
their dependants need not be updated, but the samples being reused
might have a different likelihood to what they had in the previous
iteration, which must be taken into account in the acceptance
probability. As such, each \code{assume} becomes a barrier of sorts,
segmenting the portions of the graph that need updating with each
redraw.

\begin{figure}
  \begin{minipage}{\linewidth}
    \begin{lstlisting}[language=MCore,autogobble=true]
      let example = lam inputs: [Float]. lam st.
        match pure (Gaussian 0.0 10.0) st with (st, d1) in$\label{line:running-example:prior-start}$
        match assume d1 st with (st, mean) in
        match pure (InverseGamma 5.0 125.0) st with (st, d2) in
        match assume d2 st with (st, variance) in$\label{line:running-example:prior-end}$
        match map (lam m. lam v. Gaussian m v) mean st with (st, x) in$\label{line:running-example:gaussian-start}$
        match apply x variance st with (st, d) in$\label{line:running-example:gaussian-end}$
        let f = lam st. lam input.
          match assume d st with (st, center) in$\label{line:running-example:gaussian-draw}$
          match map (lam c. logObserve input (Gaussian c 1.0)) center with (st, w) in$\label{line:running-example:observe-start}$
          weight w st in$\label{line:running-example:observe-end}$
        let st = foldl f st inputs in$\label{line:running-example:for-each}$
        (st, mean)
    \end{lstlisting}
  \end{minipage}
  \caption{An example of a (somewhat contrived) Bayesian network expressed with our approach.\label{fig:running-example:code}}
\end{figure}

\begin{figure}
    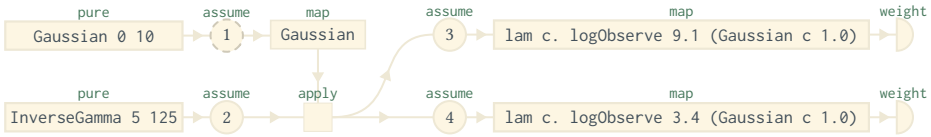

  \caption{The graph obtained from running the code in \fig{fig:running-example:code} with \code{[9.1, 3.4]} as input. Each node is annotated with its kind, and the output node is surrounded by dashes. The \code{assume} nodes in the figure are numbered, to alleviate referencing them in text.\label{fig:running-example:graph}}
\end{figure}

We now return to \fig{fig:interface} and the final two operations:
\code{join} and \code{sub}. These allow us to express more than
Bayesian networks; we can describe models that require
universality. Intuitively we do this by nesting subgraphs, where
redraws in outer graphs may discard and replace subgraphs as
necessary.

\begin{figure}
  \begin{minipage}{0.48\linewidth}
    \begin{lstlisting}[language=MCore,autogobble=true]
      let geometric = lam p. lam st.
        match pure (Bernoulli p) st $\label{line:geometric:draw-start}$
          with (st, d) in
        match assume d st with (st, c) in $\label{line:geometric:draw-end}$
        let f = lam c. lam st. if c
          then geometric p st $\label{line:geometric:recursion}$
          else pure 0 st in
        match map f c st with (st, x) in $\label{line:geometric:map}$
        match sub x st with (st, x) in
        match join x st with (st, x) in $\label{line:geometric:join}$
        map (addi 1) x st
    \end{lstlisting}
  \end{minipage}
  \hfill
  \raisebox{-.5\height}{%
    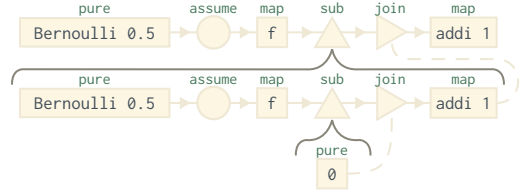
}
  \caption{An implementation of a geometric distribution (left) and the graph produced by running with \code{p = 0.5} and drawing \code{true} twice (right).\label{fig:geometric}}
\end{figure}

For example, consider the manual implementation of a geometric
distribution in \fig{fig:geometric}. The \code{geometric} function is
recursive, drawing a new random boolean
(lines~\ref{line:geometric:draw-start}-\ref{line:geometric:draw-end})
in each call and then recursing if it is \code{true}
(line~\ref{line:geometric:recursion}). Note the sequence of
\code{map}, \code{sub}, and \code{join} on
lines~\ref{line:geometric:map}-\ref{line:geometric:join}. These apply
the local function \code{f}, which selects the branch to take, then
uses \code{sub} to build the subgraph, then finally \code{join} to
extract a value from the subgraph.

Note how the model, the code on the left of \fig{fig:geometric}, is
cyclic---\code{geometric} refers to itself---but the graph remains
acyclic. This is because the graph is more akin to an execution trace
of a program than a program itself; it has one \code{sub} recording
the path taken for each probabilistic branch. As such, the graph
represents a \emph{sample} from the model, from which it is easy to
construct a new sample, \emph{not} the model itself.

With the graph-building interface now described, we can consider its
implementation.

\sectionBreakMaybe
\section{An Implementation Strategy Using Mutability}\label{sec:implementation}

Our implementation centers around two observations:

\begin{enumerate}
\item The interface ensures that nodes are created in a topologically
  sorted order (dependencies before dependents), which is a sensible
  order in which to update nodes.
\item When examining the direct dependencies of a node two things
  matter: the current value of the dependency, and if it has been
  updated since last time it was examined.
\end{enumerate}

As such, each node (except \code{weight}, which has no observable
output) has two externally visible pieces of state (a value and a
change-id) as well as an update-function represented as a
closure. Some nodes also have internal state only accessible to their
respective closure. The change-id in each node corresponds to the most
recent MCMC-iteration during which it was changed.

The update functions are accumulated in the \code{State} parameter
mentioned in \fig{fig:interface}, in the order they are created. Each
update function takes some state as input, primarily 1) the change-id
of the current MCMC iteration, 2) the weight of the current proposed
sample and 3) a list of functions to run if the proposal is
rejected. The implementation of most nodes follow the same structure,
exemplified by \code{map} below (\code{ref} creates a mutable cell,
\code{deref} and \code{modref} read from and write to such a cell,
respectively):

\begin{lstlisting}[language=MCore,autogobble=true]
  let map = lam f. lam input. lam state.
    let here = {value = ref (f (deref input.value)), id = ref 0} in$\label{line:map:init-node}$
    let update = lam u.$\label{line:map:update-start}$
      if eqi u.id (deref input.id) then$\label{line:map:dependency-check}$
        let prevValue = deref here.value in$\label{line:map:previous-value}$
        modref here.value (f (deref input.value));$\label{line:map:update-value}$
        modref here.id u.id;$\label{line:map:update-id}$
        modref u.reset (snoc (deref u.reset) (lam. modref here.value prevValue))$\label{line:map:update-reset}$
      else () in$\label{line:map:update-end}$
    ({state with updates = snoc state.updates update}, here)$\label{line:map:return}$
\end{lstlisting}

Here we first create the initial state of the node
(line~\ref{line:map:init-node}), then its \code{update} function
(lines~\ref{line:map:update-start}-\ref{line:map:update-end}), which
we store and return in the accumulated state
(line~\ref{line:map:return}). The \code{update} function itself first
checks if the input has changed
(line~\ref{line:map:dependency-check}), and if so, computes a new
\code{value} (line~\ref{line:map:update-value}). Note that each time
we change \code{value} we also update the change-id
(line~\ref{line:map:update-id}) and add a \code{reset} function for
restoring the previous \code{value}
(lines~\ref{line:map:previous-value} and~\ref{line:map:update-reset}).

Most nodes are straightforward variations of the above, but three bear
closer inspection: \code{sub}, which builds subgraphs, and the two
probabilistic nodes, \code{weight} and \code{assume}.

\code{sub} stores the update functions of the current subgraph as
extra internal state. If the input changes we rebuild the subgraph and
overwrite this state, otherwise we update the subgraph by calling each
subgraph update function in turn.

\code{weight} has no output, and thus no \code{value} or
\code{id}. However, it does have an \code{update} function that
unconditionally contributes its input as a factor in the weight of the
current sample.

\code{assume} is the only node capable of initiating change, and the
most important from an inference perspective, and thus also the most
complicated ($\sim$50 lines, compared to \ref{line:map:return}~lines
for \code{map} above). Broadly speaking, its \code{update} function
has three cases:

\begin{enumerate}
\item\label{case:redraw} \emph{If inference has picked this random
variable} to be redrawn we will update \code{value} with a new random
  sample, and also update the Metropolis-Hastings ratio used to accept
  or reject the proposal.
\item\label{case:changed-distribution} \emph{If the input has
changed,} i.e., the distribution of the random variable has changed,
  then we update the Metropolis-Hastings ratio, but leave \code{value}
  unchanged.
\item \emph{Otherwise} we do nothing.
\end{enumerate}

Additionally, the redraw can be done via a drift kernel (i.e., a
distribution parameterized by the previously drawn value) or directly
from the input distribution. We also handle an edge-case in
case~\ref{case:changed-distribution} above: if the the distribution
has changed such that the previous \code{value} has no support, then
we fall back to case~\ref{case:redraw}, drawing a new value from the
newly changed distribution.

An important consequence of this implementation is that there is some
amount of overhead for each node: both in terms of storing an
additional update function, but also updating and restoring more
mutable state. All else being equal, we would thus rather have fewer
nodes, which is a concern for the automatic translation, which we
cover in the upcoming section.

\sectionBreakMaybe
\section{An Automatic Translation}\label{sec:automatic-translation}

The interface presented in Section~\ref{sec:interface} is conceptually
clean and simple, but writing code using it can be somewhat tedious,
especially for programmers who are not used to and comfortable with
monads and related concepts in functional programming. As such, we
also provide an automatic translation from a surface language
(TreePPL) that does not distinguish between deterministic and
probabilistic values. This section outlines the basic approach and the
core ideas behind it.

Note that while this section uses some semi-formal notation, we do not
present the complete transformation, only the intuition and the most
interesting aspects.

We split the transformation into two steps: one that ignores the
\code{S} type in \fig{fig:interface} and inserts all operations except
\code{sub} (the \emph{idealized} transformation) and one that takes
\code{S} into account and inserts \code{sub} operations (the
\emph{state} transformation).

\subsection{The Idealized Transformation}

To simplify the task we assume that we always have static access to
the body of any referenced function. This implies the absence of first
class functions, which can be achieved through defunctionalization,
e.g., via lambda set
specialization~\cite{brandonBetterDefunctionalizationLambda2023}.

\paragraph{The core.}
The objective of this first transformation is to preserve the
well-typedness of an expression when the type of \code{assume} and
\code{weight} change from \code{all a. Dist a -> a} and \code{Float ->
  ()} to \code{all a. P (Dist a) -> P a} and \code{P Float -> ()}.
The transformation can be seen as a function with four inputs, and
three outputs:

\[
  \Gamma; \Lambda; R \vdash e \Rightarrow e' : \tau ; R'
\]

We first explain $\Gamma$, $e$, $e'$, and $\tau$. $\Gamma$ denotes the
values in scope (containing identifiers and their probabilistic types,
e.g., $x : \tau$) while $e$ denotes the expression to be
transformed. The primary output is $e'$, which is mostly identical to
$e$ except for the addition of calls to \code{pure}, \code{map},
\code{apply}, and \code{join}. Similarly, the output type $\tau$ is
the same as the type of $e$, except for the addition of zero or more
\code{P}s. For example:

\[
\begin{split}
  &\code{x} : \code{P Int}; \emptyset; \emptyset \vdash \code{(addi x 1, true)} \\
  \Rightarrow\ & \code{map (lam x2. (addi x2 1, true)) x} : \code{(P Int, Bool)}\ [\emptyset]
\end{split}
\]

The original expression has type \code{(Int, Bool)}, while the
transformed expression is partially probabilistic, with type \code{(P
  Int, Bool)}.

\paragraph{Supporting functions.}
$\Lambda$, $R$, and $R'$ are used to support functions. We handle
functions by monomorphization; we generate a copy for each unique set
of probabilistic input types. $\Lambda$ tracks function definitions in
scope (it contains function identifiers and their bodies, e.g., $x =
e$), while $R$ and $R'$ carry currently used monomorphizations
(function identifiers, argument types, and resulting return type,
e.g., $x\ \overline{\tau} : \tau$). For example, \code{inc} below is
duplicated because it is used with a deterministic argument as well as
a probabilistic argument:

\begin{center}
  \begin{minipage}{0.39\linewidth}
    \begin{center}
      Before
      \begin{lstlisting}[language=MCore,autogobble=true]
        let inc = lam x. addi x 1 in

        let dist = Geometric 0.5 in
        ( inc 1
        , inc (assume dist)
        )
      \end{lstlisting}
    \end{center}
  \end{minipage}
  \hfill
  \begin{minipage}{0.59\linewidth}
    \begin{center}
      After
      \begin{lstlisting}[language=MCore,autogobble=true]
        let inc1 = lam x. addi x 1 in
        let inc2 = lam x. map (lam x. addi x 1) x in
        let dist = Geometric 0.5 in
        ( inc1 1
        , inc2 (assume (pure dist))
        )
      \end{lstlisting}
    \end{center}
  \end{minipage}
\end{center}

Note that we determine which monomorphization to use at the point of
its invocation, either by finding it in $R$, or by recursively
transforming the function definition. This will not terminate for
recursive functions. As such, when we encounter the invocation of a
recursive function (with a previously unseen set of input types), we
guess that it will return a fully pure value---and record this guess
in $R$---then transform its body. If the body transforms to a
different probabilistic type we update the guess in $R$ and transform
the body again, and repeat the process until fixpoint.

This is unsatisfying from a theoretical viewpoint, and a prime
candidate for future improvement, but sufficient in practice; even
large and complicated probabilistic models are typically \emph{quite}
small when compared to programs in general. Furthermore, the process
is guaranteed to terminate, because each step can only increase how
probabilistic the return type is, and there is a maximally
probabilistic type (namely the one where \code{P} wraps the entire
type).

\paragraph{Supporting \code{match}.}
Our language has a single branching construct, \code{match}, which
checks a value against a pattern, then evaluates one branch if it fits
and another if it does not. Properly supporting this construct is more
complicated than it may first appear.

First, our type system demands that both branches produce values of
the same type. This was true before the idealized transformation, but
might not be after; one branch might be more probabilistic than the
other. Fortunately, we can always ``pretend'' a value is more
probabilistic than it actually is through the \code{Applicative} part
of \fig{fig:interface}, e.g., through \code{pure} in the simplest
case. We thus compute a least upper bound of the two branch types and
generate conversion code for both branches. For example, assuming
\code{c : Bool}, \code{l : (P Int, Bool)}, and \code{r : (Int, P
  Bool)}:

\begin{center}
  \begin{minipage}{0.39\linewidth}
    \begin{center}
      Before
      \begin{lstlisting}[language=MCore,autogobble=true]
        match c with true then
          l  -- : (P Int, Bool)
        else
          r  -- : (Int, P Bool)
      \end{lstlisting}
    \end{center}
  \end{minipage}
  \hfill
  \begin{minipage}{0.59\linewidth}
    \begin{center}
      After
      \begin{lstlisting}[language=MCore,autogobble=true]
        match c with true then
          (l.0, pure l.1)  -- : (P Int, P Bool)
        else
          (pure r.0, r.1)  -- : (P Int, P Bool)
      \end{lstlisting}
    \end{center}
  \end{minipage}
\end{center}

Note that, as above, the least upper bound might be different from
both types, requiring conversion code for both branches.

Second, the matched value might be probabilistic, but \code{match}
requires a deterministic value. This means that we may need to move
the entire \code{match} into a \code{map}. However, if the branches
return a probabilistic value, then the result of this \code{map} will
contain nested \code{P} types, which we are avoiding. In such a case
we generate conversion code for each branch such that the \code{P}
type wraps the entire type, then add a \code{join} around the entire
\code{map}. For example, assuming \code{c : P Bool} but \code{l} and
\code{r} have types as before:

\begin{center}
  \begin{minipage}{0.37\linewidth}
    \begin{center}
      Before
      \begin{lstlisting}[language=MCore,autogobble=true]
        match c with true then
          l  -- (P Int, Bool)
        else
          r  -- (Int, P Bool)
      \end{lstlisting}
    \end{center}
  \end{minipage}
  \hfill
  \begin{minipage}{0.61\linewidth}
    \begin{center}
      After
      \begin{lstlisting}[language=MCore,autogobble=true]
        let f = lam c1.
          match c1 with true then
            map (lam x. (x, l.1)) l.0 -- P (Int, Bool)
          else
            map (lam x. (r.0, x)) r.1 -- P (Int, Bool)
        in join (map f c)
      \end{lstlisting}
    \end{center}
  \end{minipage}
\end{center}

Finally, the match scrutinee may occur again in the \code{match}
branches. In such a case we reuse the unwrapped value bound by the
\code{map} lambda:

\begin{center}
  \begin{minipage}{0.225\linewidth}
    \begin{center}
      Before
      \begin{lstlisting}[language=MCore,autogobble=true]
        match i with 0
        then addi 1 i
        else i
      \end{lstlisting}
    \end{center}
  \end{minipage}
  \hfill
  \begin{minipage}{0.37\linewidth}
    \begin{center}
      Naive
      \begin{lstlisting}[language=MCore,autogobble=true]
        join (map
          (lam i1. match i1 with 0
            then map (addi 1) i
            else i)
          i)
      \end{lstlisting}
    \end{center}
  \end{minipage}
  \hfill
  \begin{minipage}{0.37\linewidth}
    \begin{center}
      After
      \begin{lstlisting}[language=MCore,autogobble=true]
        map
          (lam i1. match i1 with 0
            then addi 1 i1
            else i1)
          i
      \end{lstlisting}
    \end{center}
  \end{minipage}
\end{center}

Note that the naive translation uses \code{i} in the \code{match}
rather than \code{i1}, even though both represent the same variable,
and the latter can be used without additional graph operations.

\paragraph{Smart constructors and code generation.}
The idealized transformation makes heavy use of smart constructors, in
an effort to minimize redundant graph nodes, as implied in the
previous section. These smart constructors apply rewrites based on
laws for \code{Functor}, \code{Applicative}, and \code{Monad}. For
example, if the \code{apply} smart constructor is applied to terms
representing $\code{pure}\ e$ and $\code{pure}\ e'$ for some
expressions $e$ and $e'$, it instead applies the smart constructor for
\code{map} on $e$ and $\code{pure}\ e'$, which in turn creates the
term $\code{pure} \ (e\ e')$, creating one node instead of three.

\subsection{The State Transformation}

The second transformation step takes the \code{S} type into account,
introducing a state value to be threaded throughout the
program. Similarly to the previous step, the task is to preserve
well-typedness in the face of more precise types---in this case
re-adding \code{S}---with the additional constraint that the
\code{State} type should be used linearly; each value of this type
must be used exactly once.

Recall that each operation in \fig{fig:interface} returns a value of
type \code{S a} for some type \code{a}. Expanding the type alias
\code{S} yields \code{State -> (State, a)}, i.e., each operation takes
and returns a \code{State} value in addition to its other inputs and
return value. Each such operation will consume the previously active
\code{State} value and produce a new one. The state transformation is
responsible for threading these values from one operation to the next,
which is a largely mechancial effort, with only two language
constructs deserving of further scrutiny.

First, the \code{State} value is passed to user-defined functions
through an extra argument if the function contains graph operations,
and not passed at all otherwise.

Second, depending on the initial input program, some graph operations
may end up used inside the lambda of a \code{map} operation, namely if
a \code{match} examines a probabilistic value (c.f. \emph{supporting
\code{match}} in the previous subsection). This is an issue for
linearity, because said lambda may run any number of times during
inference, thus no external \code{State} can be used. We solve this by
adding a \code{State} parameter to the lambda, then adding a
\code{sub} operation to discharge it. For example, one of the examples
of the previous subsection is further transformed as follows:

\begin{center}
  \begin{minipage}[t]{0.46\linewidth}
    \begin{center}
      Before
      \begin{lstlisting}[language=MCore,autogobble=true]
        let program =
          let f = lam c1.
            match c1 with true then
              map (lam x. (x, l.1)) l.0
            else
              map (lam x. (r.0, x)) r.1
          in join (map f c)
      \end{lstlisting}
    \end{center}
  \end{minipage}
  \hfill
  \begin{minipage}[t]{0.51\linewidth}
    \begin{center}
      After
      \begin{lstlisting}[language=MCore,autogobble=true]
        let program = lam st1.
          let f = lam c1. lam st2.
            match c1 with true then
              map (lam x. (x, l.1)) st2 l.0
            else
              map (lam x. (r.0, x)) st2 r.1 in
          match map f st1 with (st3, x) in
          match sub x st3 with (st4, x) in
          join x st4
      \end{lstlisting}
    \end{center}
  \end{minipage}
\end{center}

Note the additional argument to the program as a whole, the passing of
\code{stN} variables, and the new call to \code{sub}.

This use of \code{sub} is what gives rise to sub-graphs. Running the
function encompassing the entire program produces the main graph, the
top layer whose topology will remain constant throughout inference,
while each update flowing into a \code{sub} node rebuilds a sub-graph
by running the underlying \code{State -> (State, a)} function.

\sectionBreakMaybe
\section{Conclusion and Future Work}

In this paper we connect two disparate topics, functional reactive
programming and probabilistic programming, and use the combination to
do probabilistic inference according to an MCMC inference method. Each
program produces a graph representing a sample, which can be modified
to produce another sample, while recomputing only the portions of the
graph that have changed. The goal of our approach in runtime
efficiency, and preliminary results appear promising.

There are several important avenues of future work. First, as
mentioned above, our approach would benefit from empirical validation,
i.e., running inference on larger models and datasets. Second, there
are a number of common precision-increasing approaches that could
benefit from a direct integration into our graph, e.g.,
Rao-Blackwellization~\cite{blackwellConditionalExpectationUnbiased1947},
pruning~\cite{felsensteinMaximumLikelihoodMinimumSteps1973}, or
delayed sampling~\cite{murrayDelayedSamplingAutomatic2018}. Third,
though our approach is optimized for MCMC inference, the core
abstraction should also be usable for more exotic inference approaches
that, e.g., mix MCMC and SMC.

\begin{acks}
This work was supported by the Swedish Research Council (grant
2021-04830 to FR) and by the Knut and Alice Wallenberg Foundation
through the DarkTree project (grant 2024.0076 to FR and DB). The
authors would like to thank Erik Danielsson and Thimothée Virgoulay
for discussion and feedback.
\end{acks}

\printbibliography

\end{document}